\documentclass{INTERSPEECH2023}
\usepackage{multirow}
\usepackage{graphicx}
\usepackage{url}
\usepackage{enumitem}
\usepackage{threeparttable}
\usepackage{hyperref}

\interspeechcameraready

\title{P-vectors:A Parallel-Coupled TDNN/Transformer Network for \\ Speaker Verification}
\name{Xiyuan Wang$^{1\ast}$, Fangyuan Wang$^{1\ast}$, Bo Xu$^{1,2,3,\dag}$, Liang Xu$^4$,Jing Xiao$^5$ \thanks{$^{\ast}$Equal contribution, $^{\dag}$corresponding author} \thanks{$^1$Our codes are available at \url{https://github.com/xyw7/}}}
\address{
  $^1$Institute of Automation, Chinese Academy of Sciences\\
  $^2$School of Future Technology, $^3$University of Chinese Academy of Sciences\\
  $^4$GammaLab, PingAn OneConnect, $^5$PingAn Insurance (Group) Company of China}
\email{xiyuan.wang@ia.ac.cn, fangyuan.wang@ia.ac.cn, xubo@ia.ac.cn, xlpaul@126.com, xiaojing661@pingan.com.cn}

\begin{document}

\maketitle
 
\begin{abstract}
Typically, the Time-Delay Neural Network (TDNN) and Transformer can serve as a backbone for Speaker Verification (SV).  
Both of them have advantages and disadvantages from the perspective of global and local feature modeling. How to effectively integrate these two style features is still an open issue.
In this paper, we explore a Parallel-coupled TDNN/Transformer Network (p-vectors) to replace the serial hybrid networks.
The p-vectors allows TDNN and Transformer to learn the complementary information from each other through Soft Feature Alignment Interaction (SFAI) under the premise of preserving local and global features.
Also, p-vectors uses the Spatial Frequency-channel Attention (SFA) to enhance the spatial interdependence modeling for input features. 
Finally, the outputs of dual branches of p-vectors are combined by Embedding Aggregation Layer (EAL).
Experiments$^1$ show that p-vectors outperforms MACCIF-TDNN and MFA-Conformer with relative improvements of 11.5$\%$ and 13.9$\%$ in EER on VoxCeleb1-O.
\end{abstract}

\section{Introduction}

The Time Delay Neural Network (TDNN)~\cite{tdnn1} based systems~\cite{tdnn2,ecapa} have recently proved their superiority over the conventional i-vectors~\cite{Dehak2011FrontEndFA,ivector2} in speaker verification (SV) task. 
On the other hand, several works~\cite{svector,finegrain2,transfomer_1,maccif,mfconformer} have tried to introduce the Transformer~\cite{transformer} and its variants~\cite{conformer_real} to build speaker embedding, seeing that self-attention has shown outstanding performances in a wide range of speech-processing tasks~\cite{transinspeech}.

There are mainly two types of methods to utilize Transformer in SV systems, the solely Transformer based models~\cite{svector,finegrain2,transfomer_1} and the hybrid ones~\cite{maccif,mfconformer,hybrid3}.
The typical solely Transformer based model s-vector~\cite{svector} replaces TDNN modules with Transformer encoder layers in the x-vector framework to capture the global features. However, these methods typically suffer notable degradation as they ignore the fine-grained local features that have been proven beneficial to the overall performance~\cite{finegrain2,finegrain1}.
The hybrid methods try to find a way to combine the local features captured by the convolutional operators and the global features built by the multi-heads self-attention (MHSA) mechanism. MACCIF-TDNN~\cite{maccif} proposed a Transformer and ECAPA-TDNN~\cite{ecapa} hybrid network, where Transformer is connected in series after ECAPA-TDNN to fuse local features into global representation. 
MFA-Conformer~\cite{mfconformer} introduces the Conformer, which has demonstrated its power in many ASR tasks, to the SV task. 
\begin{figure}[ht]
\centering
\includegraphics[width=0.83\linewidth]{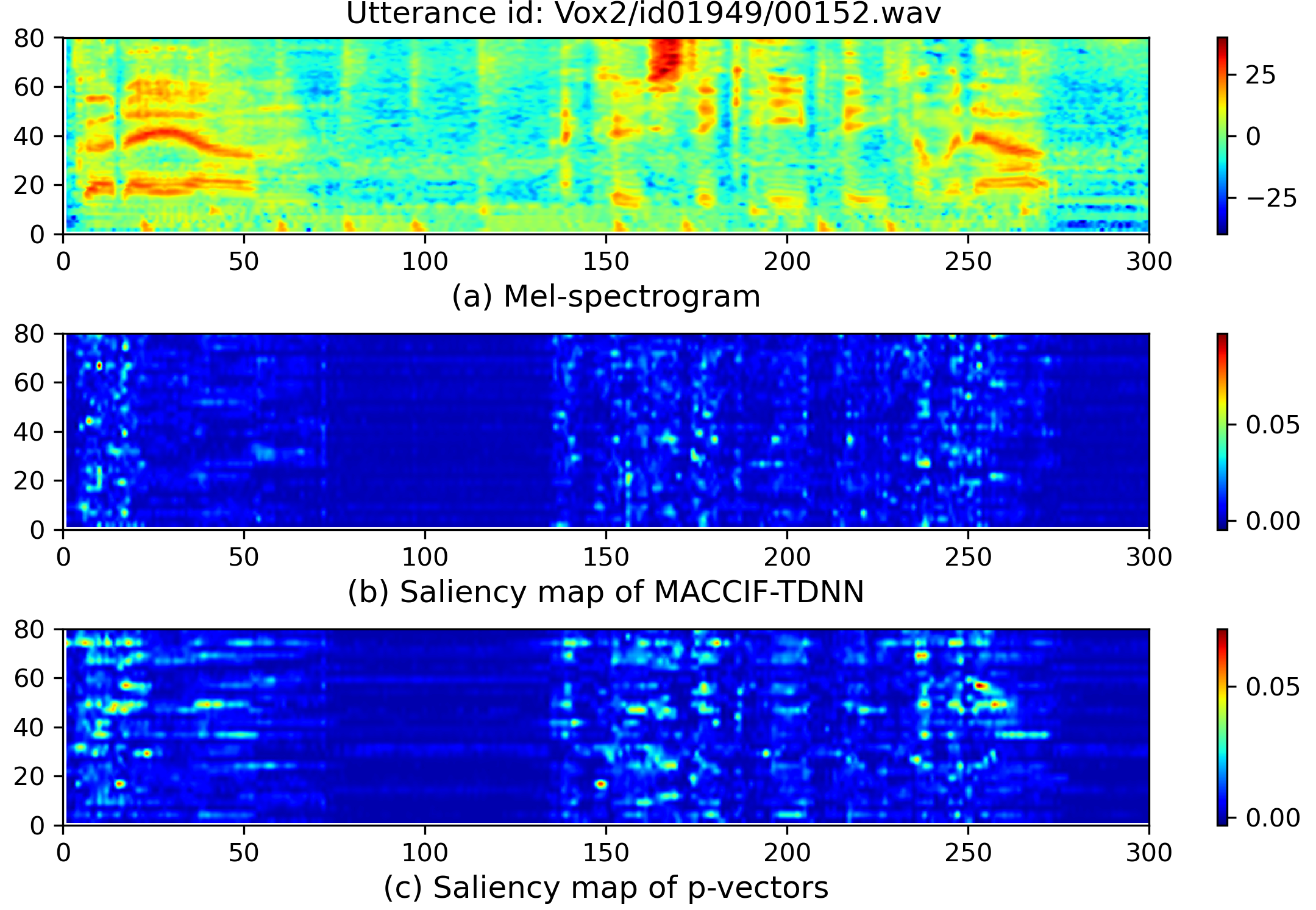}
\caption{(a) The Mel-spectrogram of the example utterance. (b) and (c) are saliency maps obtained by applying the Layer-CAM~\cite{layercam,layercam2} to the same convolutional layer (the second Conv1d layer in the third SE-Res2Block) of MACCIF-TDNN~\cite{maccif} and p-vectors, respectively. The brighter the color, the more important the region to identify the corresponding speaker~\cite{layercam2}.}
\label{fig0}
\end{figure}
The core component of ~\cite{mfconformer} is a macaron-like structure serially connected by the feed-forward module, MHSA module, and convolution module (Conformer Block). 
\cite{mfconformer} integrates local and global features by multiple cascaded Conformer blocks and aggregates them to obtain the speaker embedding.
Both methods are variants of the serial concatenation of convolutional layers and MHSA modules.
However, since the convolutional operation and the MHSA mechanism focus on different modeling perspectives, the features extracted from these two styles naturally have information misalignment and semantic divergence~\cite{conformer}. Therefore, the serial stacking of these two feature modeling operations may distort the feature stream and force it to fit different modeling mechanisms, resulting in the degradation of the extracted local or global features stored in the feature stream and the inefficient interaction between them. 
To avoid, we adopt a parallel structure with Soft Feature Alignment Interaction (SFAI) to couple them (p-vectors). 
On the premise of preserving local and global features, p-vectors can align the two styles of information and allow them to interact selectively.
Moreover, to further fuse the local and global information, we propose the Embedding Aggregation Layer (EAL) to aggregate the highly differentiated embeddings extracted from the both styles and generate the final speaker embedding.
Figure~\ref{fig0} shows that the saliency map of p-vectors has more horizontal bright areas than MACCIF-TDNN's, indicating that the Conv1d layer trained under the SFAI can borrow more long-range feature from the MHSA mechanism than the serial MACCIF-TDNN in modeling.
\begin{figure*}[ht]
\centering
\includegraphics[width=0.9\textwidth]{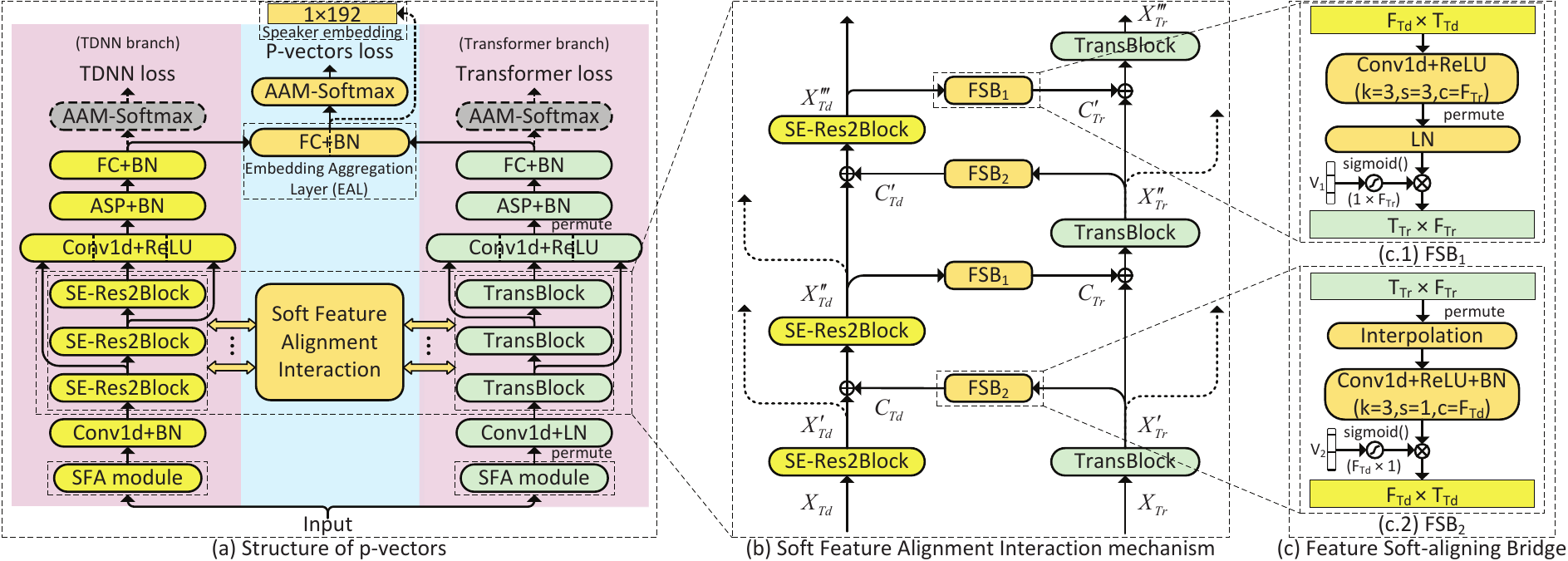}
\caption{(a) The structure of p-vectors. Note that the TDNN loss, Transformer loss and the classifiers (the gray dashed boxes) are used in the first training stage (mentioned in section~\ref{sec:trainstrategy}). (b) Soft feature alignment interaction mechanism. Note that the dashed arrows extending from the TDNN and Transformer branch are used for aggregation (c) Feature soft-aligning bridge. $V_{1}$ and $V_{2}$ are trainable vectors ($F_{Td}$, $T_{Td}$, $F_{Tr}$ and $T_{Tr}$ correspond to the frequency and temporal length of the TDNN and Transformer feature, respectively).}
\label{fig1}
\end{figure*}

Besides, as a feature augmentation, the attention mechanism can further improve the performance of SV.
MFA~\cite{mfa} emphasizes important local frequency region by applying squeeze-and-excitation (SE)~\cite{se} to a flattened frequency-channel map.
However, the flattening process may destroy the instance-agnostic information hidden in the spatial relationship between frequency bins and channels. 
To improve, we propose a spatial frequency-channel attention (SFA) to adaptively re-scale local feature responses by modeling the interdependence between channels and frequency bins from the spatial perspective.

Our contributions can be summarized as follows:
\begin{itemize}
\item We propose a TDNN-Transformer parallel-coupled structure with the SFAI mechanism and the EAL to effectively integrate the local features and global representations.
\item We propose a spatial frequency-channel attention to enhance the feature representation ability from the spatial perspective.
\end{itemize}

We conduct numerous experiments and the results show that p-vectors can integrate local and global information more effectively even with fewer parameters. Compared to serial hybrid models, p-vectors outperforms MACCIF-TDNN~\cite{maccif} and MFA-Conformer~\cite{mfconformer} with relative improvements of 11.5$\%$ and 13.9$\%$ in EER(\%) when trained on VoxCeleb2 development set.

\section{Proposed p-vectors}
\subsection{Structure overview}
As Figure~\ref{fig1}(a) shows, p-vectors consists of: (1) the SFA modules; (2) the TDNN-Transformer dual branches; (3) the SFAI mechanism; (4) the EAL.
In terms of functionality, the dual branches with the SFA modules (pink background) extract features rich in local/global features while the SFAI mechanism and the EAL (blue background) focus on feature interaction and aggregation.
The yellow, green, and orange boxes in Figure~\ref{fig1}(a) correspond to modules/features of the TDNN, the Transformer branch, and feature interaction/aggregation, respectively.
\subsection{Spatial frequency-channel attention}
Recently, various attention mechanisms have been applied to SV~\cite{ecapa,att1,att2,mfa} and have proven their effectiveness.
~\cite{mfa} proposed the frequency-channel attention by applying SE~\cite{se} to the flattened frequency-channel map to emphasize the important channels and frequency. 
However, some spatial dependencies between frequency bins and channels may be lost during the flattening process. 
Therefore, we propose the spatial frequency-channel attention (SFA) to exploit the spatial relationship between frequency bins and channels. 
\begin{figure}[h]
\centering
\includegraphics[width=0.68\linewidth]{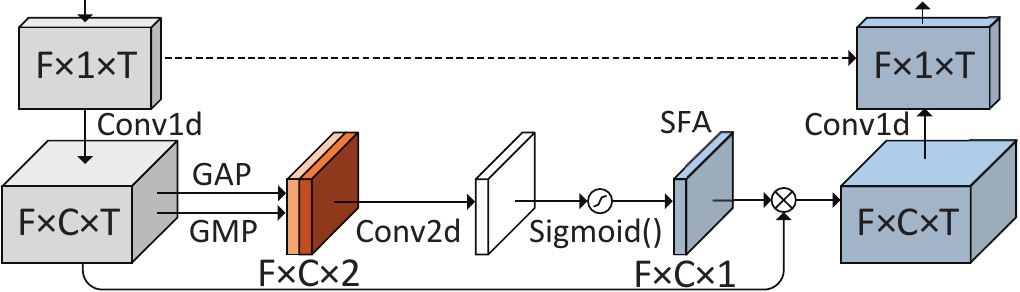}
\caption{Structure of the proposed SFA}
\label{fig2}
\end{figure}
As Figure~\ref{fig2} shows, the channels of the input acoustic feature are first expanded by using a Conv1d layer to enrich the representation. 
Next, Global Average Pooling (GAP) and Global Maximum Pooling (GMP) are applied to the expanded feature to generate the frequency channel maps that contain the basic and unique characteristics of the speaker.
Then a Conv2d layer is applied to the concatenated frequency-channel map to generate the SFA, which contains the spatial information of the frequency-channel.
Finally, the expanded feature is multiplied by the SFA processed by the sigmoid function, and the number of channels is reduced by the Conv1d layer to obtain the SFA-emphasized feature.
\subsection{Dual branches}
As the left part of Figure~\ref{fig1}(a) shows, the TDNN branch consists of an SFA module and the ECAPA-TDNN~\cite{ecapa}. As the core component of~\cite{ecapa}, the SE-Res2Block integrates TDNN-specific SE-blocks~\cite{se} with hierarchically residual connected Res2Net~\cite{res2net} modules to extract local features from the frame level.
As the right part of Figure~\ref{fig1}(a) shows, we replace the SE-Res2Blocks in the TDNN branch with the TransBlocks which consists of three concatenated Transformer encoder layers. The MHSA-based TransBlocks over the unrestricted context can capture the global representations from the utterance level. 
\begin{table*}[t]
\centering
\caption{Performance overview of all systems on VoxCeleb1 and VoxSRC22 validation set}
\label{tab:table1}
\begin{threeparttable}
\begin{tabular}{ccccccccccc}
\hline
\makebox[0.07\textwidth]{\multirow{2}{*}{\textbf{Architecture}}} &
  \makebox[0.05\textwidth]{\multirow{2}{*}{\textbf{Channels}}} &
  \makebox[0.06\textwidth]{\multirow{2}{*}{\textbf{\#Params}}} &
  \multicolumn{2}{c}{\textbf{VoxCeleb1-O}} &
  \multicolumn{2}{c}{\textbf{VoxCeleb1-E}} &
  \multicolumn{2}{c}{\textbf{VoxCeleb1-H}} &
  \multicolumn{2}{c}{\textbf{VoxSRC22-val}} \\ \cline{4-11} 
 &
   &
   &
  \makebox[0.048\textwidth]{\textbf{EER(\%)}} &
  \makebox[0.048\textwidth]{\textbf{MinDCF}} &
  \makebox[0.048\textwidth]{\textbf{EER(\%)}} &
  \makebox[0.048\textwidth]{\textbf{MinDCF}} &
  \makebox[0.048\textwidth]{\textbf{EER(\%)}} &
  \makebox[0.048\textwidth]{\textbf{MinDCF}} &
  \makebox[0.048\textwidth]{\textbf{EER(\%)}} &
  \makebox[0.048\textwidth]{\textbf{MinDCF}} \\ \hline
ECAPA-TDNN~\cite{ecapa} &
  512 &
  6.2M &
  1.01 &
  0.1274 &
  1.24 &
  0.1418 &
  2.32 &
  0.2181 &
  - &
  - \\
ECAPA-TDNN~\cite{ecapa}(Repro) &
  512 &
  7.3M &
  1.0050 &
  0.1449 &
  1.2390 &
  0.1300 &
  2.4258 &
  0.2335 &
  3.4266 &
  0.3647 \\
MACCIF-TDNN~\cite{maccif} &
  512 &
  - &
  1.19 &
  0.148 &
  1.47 &
  0.158 &
  2.48 &
  0.235 &
  - &
  - \\
MACCIF-TDNN~\cite{maccif}(Repro) &
  512 &
  15.3M &
  0.9678 &
  0.1344 &
  1.2048 &
  0.1249 &
  2.3168 &
  0.2293 &
  3.2881 &
  0.3465 \\
MFA-Conformer~\cite{mfconformer}* &
  - &
  20.5M &
  0.64 &
  0.081 &
  - &
  - &
  - &
  - &
  - &
  - \\ 
MFA-Conformer~\cite{mfconformer}(Repro) &
  - &
  20.9M &
  0.9944 &
  0.1528 &
  1.2271 &
  0.1311 &
  2.3971 &
  0.2329 &
  3.3192 &
  0.3579 \\ \hline
MFA~\cite{mfa} &
  512 &
  7.3M &
  0.8561 &
  0.0913 &
  1.1382 &
  0.1208 &
  2.0487 &
  0.1897 &
  - &
  - \\ \hline
\textbf{p-vectors w/o SFA} &
  512 &
  \textbf{15.0M} &
  \textbf{0.9253} &
  \textbf{0.1221} &
  \textbf{1.1560} &
  \textbf{0.1213} &
  \textbf{2.2378} &
  \textbf{0.2187} &
  \textbf{3.2073} &
  \textbf{0.3337} \\
\textbf{p-vectors} &
  512 &
  \textbf{15.1M} &
  \textbf{0.8561} &
  \textbf{0.1199} &
  \textbf{1.1170} &
  \textbf{0.1201} &
  \textbf{2.1120} &
  \textbf{0.2081} &
  \textbf{3.0839} &
  \textbf{0.3171} \\ \hline

\end{tabular}
\begin{tablenotes}
        \footnotesize
        \item[*] This result reported in~\cite{mfconformer} is trained on both the development set of VoxCeleb1\&2 (7205 speakers), while our reproduced models, for comparison purposes, are all trained on the development set of VoxCeleb2 (5994 speakers) only.
      \end{tablenotes}
    \end{threeparttable}
\end{table*}
\subsection{Soft feature alignment interaction mechanism}
The TDNN branch gradually extracts local features through frame-level Conv1d layers while the Transformer branch aggregates global speaker characteristics with the MHSA mechanism. 
To interact the local and global features, the dual branches are bridged by the soft feature alignment interaction mechanism, as Figure~\ref{fig1}(b) shown.
Formally, it can be described as:
\begin{equation}
\begin{aligned}
\textbf{X}^{'}_{Td} &= SE\mbox{-}Res2Block(\textbf{X}_{Td})\\
\textbf{X}^{''}_{Td} &= SE\mbox{-}Res2Block(\textbf{X}^{'}_{Td}+\textbf{C}_{Td})\\
\textbf{X}^{'''}_{Td} &= SE\mbox{-}Res2Block(\textbf{X}^{''}_{Td}+\textbf{C}^{'}_{Td})\\
\textbf{X}^{'}_{Tr} &= TransBlock(\textbf{X}_{Tr})\\
\textbf{X}^{''}_{Tr} &= TransBlock(\textbf{X}^{'}_{Tr}+\textbf{C}_{Tr})\\
\textbf{X}^{'''}_{Tr} &= TransBlock(\textbf{X}^{''}_{Tr}+\textbf{C}^{'}_{Tr})
\end{aligned}
\end{equation}
\begin{equation}
\begin{aligned}
\textbf{C}_{Td} &= FSB_2(\textbf{X}^{'}_{Tr})\\
\textbf{C}^{'}_{Td} &= FSB_2(\textbf{X}^{''}_{Tr})\\
\textbf{C}_{Tr} &= FSB_1(\textbf{X}^{''}_{Td})\\
\textbf{C}^{'}_{Tr} &= FSB_1(\textbf{X}^{'''}_{Td})
\end{aligned}
\end{equation}
The TDNN features $[\textbf{X}_{Td},\textbf{X}^{'}_{Td},\textbf{X}^{''}_{Td},\textbf{X}^{'''}_{Td}]$ and the complementary features added to the TDNN branch $[\textbf{C}_{Td},\textbf{C}^{'}_{Td}]\in \mathbb{R}^{F_{Td} \times T_{Td}}$, the Transformer features $[\textbf{X}_{Tr},\textbf{X}^{'}_{Tr},\textbf{X}^{''}_{Tr},\textbf{X}^{'''}_{Tr}]$ and the complementary features added to the Transformer branch $[\textbf{C}_{Tr},\textbf{C}^{'}_{Tr}]\in \mathbb{R}^{T_{Tr} \times F_{Tr}}$, 
where $F_{Td}$, $T_{Td}$, $F_{Tr}$ and $T_{Tr}$ are the frequency bins and temporal dimension of the features in the TDNN and Transformer branch, respectively.

The different modeling mechanisms lead to semantic divergence and feature misalignment. To align the features of the two branches and fill the semantic gaps during interaction, we propose the feature soft-aligning bridges (${FSB}_{1}$ and ${FSB}_{2}$). 
As shown in Figure~\ref{fig1}(c.1), in the ${FSB}_{1}$, the TDNN feature first requires getting through a Conv1d layer to align the frequency bins and the temporal dimension before the dimension permuting. 
Then, to mitigate the effects of semantic divergence, it times an $F_{Tr}$-dimensional trainable vector ($V_{1}$) processed by the sigmoid function, which helps to select the useful information and filter out the unnecessary information.  
Finally, the processed TDNN feature is added to the Transformer branch.
As shown in Figure~\ref{fig1}(c.2), in the $FSB_{2}$, after the dimension permuting, the Transformer feature is upsampled to align the temporal scale. Then the frequency bins are aligned with that of the TDNN features through a Conv1d layer. Similarly, the feature is multiplied by an $F_{Td}$-dimensional trainable vector ($V_{2}$) processed by the sigmoid function and added to the TDNN branch.
Meanwhile, 1-D layer normalization (LN) and 1-D batch normalization (BN) are used to regularize the features.

\subsection{Embedding aggregation layer}
Generally, the output of the FC layer before the classifier is used as the speaker embedding in most SV task.
To further fuse the complementary local and global information, as shown in Figure~\ref{fig1}(a), we aggregate the highly differentiated speaker embeddings of the dual branches into one FC layer (the EAL) with 1-D BN to generate the final speaker embedding.

\section{Experiments setup}
\subsection{Database and data augmentation}
VoxCeleb is an audio-visual dataset extracted from interview videos. In our experiments, the VoxCeleb2~\cite{vox2} development set containing 5994 speakers is used for training. To diversify the training data, we augment the data with reverberate data (RIR dataset~\cite{rir}) and noise data (MUSAN dataset~\cite{musan}). The remaining three augmentations are generated using the open-source SoX (tempo up, tempo down) and FFmpeg libraries.

\subsection{Implementation details}
For a fair comparison, all experiments are conducted using SpeechBrain Toolkit\footnotemark[2].
A fixed length of 3-second segments are randomly sampled from the utterances. The input features are 80-dimensional Fbanks with a window length of 25 ms and a frame-shift of 10 ms. No voice activity detection is performed. All models are trained using additive margin softmax loss~\cite{aam} with a margin of 0.2 and a scaling factor of 30. 
Note that to reduce the mismatch between the reproduced baseline systems and the reference papers, we adopted their original implementation details separately, and trained each for 30 epochs.
\footnotetext[2]{\url{https://speechbrain.github.io/}}
\subsection{Training strategy}
\label{sec:trainstrategy}
To avoid the influence of premature information interaction between the two branches on the modeling and to improve the robustness, the training phase is divided into two stages. 
First, TDNN and Transformer branches are trained independently for 24 epochs to extract pure local details and global representations. 
Second, we remove their classifiers and bridge them by the feature interaction and aggregation module, which is the p-vectors. Then we transfer the independently trained weights of the two branches to p-vectors as the pre-trained weights and train p-vectors for 6 epochs.
Using the triangular2 policy~\cite{cyc}, the cyclical learning rate varies between 1e-8 and 1e-3 with the Adam optimizer. The duration of one cycle is set to 6 epochs.

\subsection{Evaluation protocol}
We evaluate the models on four trails including the VoxCeleb1-O, VoxCeleb1-E, VoxCeleb1-H~\cite{vox1}, and the VoxSRC2022~\cite{voxsrc} validation set.
Cosine distance with adaptive s-norm~\cite{asnorm} is used for scoring. We report the performance in terms of equal error rate (EER) and the minimum detection cost function (minDCF) with $P_{target}$ = 0.01 and $C_{FA}$ = $C_{Miss}$ = 1.

\section{Results}
\subsection{Parallel structure with interaction vs. serial structures}
As shown in Table~\ref{tab:table1}, compared with the serial convolution-MHSA hybrid structure ~\cite{maccif} and ~\cite{mfconformer}, our interaction-based parallel structure with fewer parameters obtains relative improvements of 4.4\%, 6.9\% in EER(\%) and 3.5\%, 15.1\% in MinDCF, respectively. The results indicate our proposed structure can combine local features and global representations more effectively, which is further demonstrated in Figure~\ref{fig0}.
After integrating the SFA module, the relative improvements achieved 11.5\%, 9.9\% in EER(\%), and 13.9\%, 20.7\% in MinDCF, respectively.
Note that our work aims to explore effective hybrid structures, the state-of-the-art model MFA~\cite{mfa} as an add-on technique can be integrated into p-vectors to further improve the performance.  
\begin{table}[h]
\centering
\caption{Ablation experiments on VoxCeleb1-O}
\label{table2}
\begin{threeparttable}
\begin{tabular}{llc}
\hline
  & \multicolumn{1}{c}{\textbf{Model}}  & \makebox[0.03\linewidth]{\textbf{EER}} \\ \hline
\makebox[0.02\linewidth]{A.1 }   & ECAPA-TDNN + FA~\cite{mfa} (Repro)& 0.98                \\ 
\makebox[0.02\linewidth]{A.2 }    & ECAPA-TDNN + SFA (TDNN branch) & 0.95                \\ \hline
\makebox[0.02\linewidth]{B.1 } & TDNN branch (solely trained)           & 0.95             \\
\makebox[0.02\linewidth]{B.2 } & Transformer branch (solely trained)          & 2.41             \\ \hline
\makebox[0.02\linewidth]{C.1*} & TDNN branch (jointly trained)               & 0.91             \\
\makebox[0.02\linewidth]{C.2*} & Transformer branch (jointly trained)            & 1.46             \\ \hline
\makebox[0.02\linewidth]{D.1*}   & dual branches + EAL            & 0.92             \\ 
\makebox[0.02\linewidth]{D.2*}  & dual branches + SFAI (w/o $V_{i}$) + EAL      & 0.89               \\ 
\makebox[0.02\linewidth]{D.3*}   & dual branches + SFAI (w/ $V_{i}$)+ EAL (\textbf{p-vectors})            & \textbf{0.86}  \\ \hline    
\end{tabular}
\begin{tablenotes}
        \footnotesize
        \item[*] following the two-stage training strategy (transfer the 24th-epoch weights from B.1\&B.2 to the bridged model and train for 6 epochs)
      \end{tablenotes}
    \end{threeparttable}
\end{table}

\subsection{Ablation experiments}

\noindent\textbf{Spatial Frequency-channel Attention (SFA)}
We reproduce the frequency-channel attention (FA)~\cite{mfa} and integrate it into ECAPA-TDNN as the experiment A.1 and compare it with the experiment A.2 which integrates the SFA into ECAPA-TDNN (same structure as TDNN branch). 
The result in Table~\ref{table2} shows that the SFA can reveal and exploit the intrinsic relationship between frequency bins and channels in spatial perspective to improve the performance of the model as compared to the FA.

\noindent\textbf{Soft Feature Alignment Interaction mechanism (SFAI)}. 
In experimental group B, we train the solely TDNN branch and Transformer branch independently for 30 epochs each.
In experimental group C, we keep the classifiers of both branches and bridge them by the SFAI. After the two-stage training, we evaluate the respective embeddings of the two branches independently.
Comparing the experimental groups B and C, Table~\ref{table2} shows that the TDNN and Transformer branches trained with the SFAI achieve relative improvements in EER(\%) of 4.21\% and 39.41\%, respectively, indicating that both branches benefit from the SFAI. In addition, we noticed that the improvement brought by the TDNN branch to the Transformer branch is huge. 
We believe this is because speaker verification/recognition is essentially a fine-grained classification task, so the local details in voiceprint is crucial. Therefore, the information from the TDNN branch can greatly compensate for the shortcomings of the Transformer branch and improve its performance.

\noindent\textbf{Embedding Aggregation Layer (EAL)}. 
In experiment D.1, we only use the EAL to bridge the two branches.
The comparison of D.1 and B.1/B.2 shows that the aggregation of the speaker embeddings obtained from the two brunches can improve the performance, which also indicates the complementarity of the speaker embeddings.
In experiment D.3, we bridge the two branches by both the SFAI and the EAL (p-vectors).
Comparing the D.3 with C.1/C.2, after aggregating the embeddings from the SFAI-bridged two branches, the performance is further improved, which also proves the effectiveness of the EAL.  
\begin{figure}[t]
\centering
\includegraphics[width=1\linewidth]{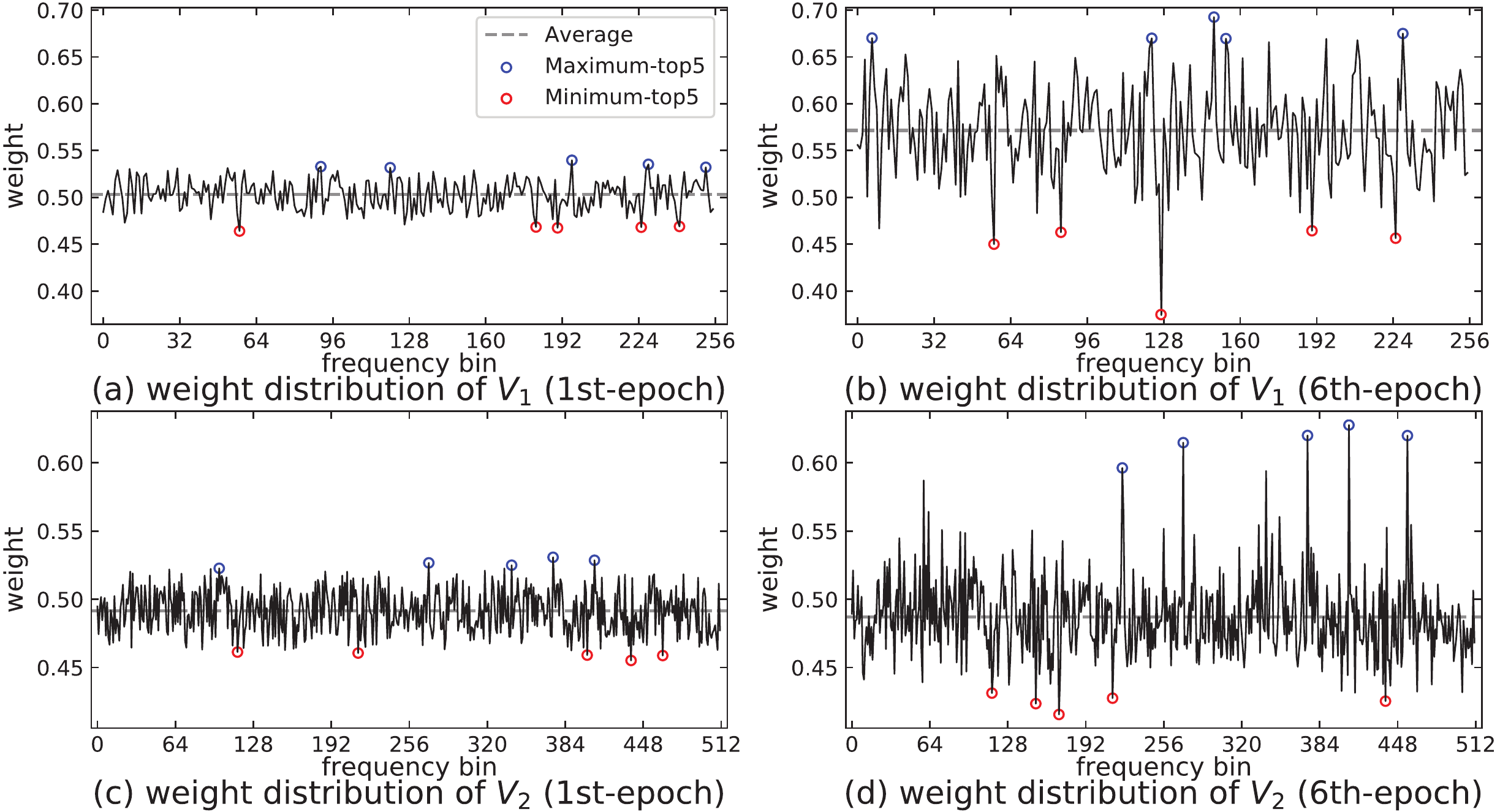}
\caption{As training progresses, the weight distribution of the trainable vectors $V_{i}$ ($V_{1}$ and $V_{2}$ of the first $FSB_{1}$ and $FSB_{2}$ in the SFAI, respectively) are gradually differentiated. The $V_{i}$ learn to selectively enhance or suppress certain frequency bins, which helps to align the information between the two bunches and mitigate the effects of semantic divergence.}
\label{fig0s}
\end{figure}

\noindent\textbf{Feature soft alignment}. 
Different from the experiment D.3, in the experiment D.2, we remove all the trainable vectors $V_{i}$ of the $FSB_{i}$ ($FSB_{1}$ and $FBS_{2}$) in the SFAI. 
Comparing the experiment D.2 and D.3, as shown in Table~\ref{table2}, without the adjustment of $V_{i}$, the performance relatively decreases by 3.37\% in EER(\%).
The result indicates that the trainable vectors help to align the information of the two branches and play a moderating role in the interaction. Figure~\ref{fig0s} can further prove this.

\section{Conclusion}
In this paper, we propose a TDNN-Transformer parallel-coupled structure which leverages the convolutional operations and the MHSA mechanism to selectively interact and aggregate the local and global information. We also propose the SFA to exploit the spatial interdependence between the frequency bins and channels. 
Evaluating results shows that the proposed p-vectors significantly outperforms the SOTA serial hybrid systems on VoxCeleb1 test sets and VoxSRC22 validation set.

\section{Acknowledgements}
This work is supported by the National Innovation 2030 Major S\&T Project of China under Grant No.2020AAA0104202 and the National Natural Science Foundation of China (62206294).

\bibliographystyle{IEEEtran}
\bibliography{mybib}

\begin{thebibliography}{10}
\providecommand{\url}[1]{#1}
\csname url@samestyle\endcsname
\providecommand{\newblock}{\relax}
\providecommand{\bibinfo}[2]{#2}
\providecommand{\BIBentrySTDinterwordspacing}{\spaceskip=0pt\relax}
\providecommand{\BIBentryALTinterwordstretchfactor}{4}
\providecommand{\BIBentryALTinterwordspacing}{\spaceskip=\fontdimen2\font plus
\BIBentryALTinterwordstretchfactor\fontdimen3\font minus
  \fontdimen4\font\relax}
\providecommand{\BIBforeignlanguage}[2]{{%
\expandafter\ifx\csname l@#1\endcsname\relax
\typeout{** WARNING: IEEEtran.bst: No hyphenation pattern has been}%
\typeout{** loaded for the language `#1'. Using the pattern for}%
\typeout{** the default language instead.}%
\else
\language=\csname l@#1\endcsname
\fi
#2}}
\providecommand{\BIBdecl}{\relax}
\BIBdecl

\bibitem{tdnn1}
A.~Waibel, T.~Hanazawa, G.~Hinton, K.~Shikano, and K.~Lang, ``Phoneme
  recognition using time-delay neural networks,'' \emph{IEEE Transactions on
  Acoustics, Speech, and Signal Processing}, vol.~37, no.~3, pp. 328--339,
  1989.

\bibitem{tdnn2}
J.~Thienpondt, B.~Desplanques, and K.~Demuynck, ``Integrating frequency
  translational invariance in tdnns and frequency positional information in 2d
  resnets to enhance speaker verification,'' in \emph{Interspeech}, 2021.

\bibitem{ecapa}
B.~Desplanques, J.~Thienpondt, and K.~Demuynck, ``Ecapa-tdnn: Emphasized
  channel attention, propagation and aggregation in tdnn based speaker
  verification,'' in \emph{INTERSPEECH}, 2020.

\bibitem{Dehak2011FrontEndFA}
N.~Dehak, P.~Kenny, R.~Dehak, P.~Dumouchel, and P.~Ouellet, ``Front-end factor
  analysis for speaker verification,'' \emph{IEEE Transactions on Audio,
  Speech, and Language Processing}, vol.~19, pp. 788--798, 2011.

\bibitem{ivector2}
D.~Snyder, D.~Garcia-Romero, D.~Povey, and S.~Khudanpur, ``{Deep Neural Network
  Embeddings for Text-Independent Speaker Verification},'' in \emph{Proc.
  Interspeech 2017}, 2017, pp. 999--1003.

\bibitem{svector}
J.~MetildaSagayaMaryN., S.~V. Katta, and S.~Umesh, ``S-vectors: Speaker
  embeddings based on transformer's encoder for text-independent speaker
  verification,'' \emph{ArXiv}, vol. abs/2008.04659, 2020.

\bibitem{finegrain2}
B.~Han, Z.~Chen, and Y.~Qian, ``Local information modeling with self-attention
  for speaker verification,'' \emph{ICASSP 2022 - 2022 IEEE International
  Conference on Acoustics, Speech and Signal Processing (ICASSP)}, pp.
  6727--6731, 2022.

\bibitem{transfomer_1}
P.~Safari, M.~India, and J.~Hernando, ``Self-attention encoding and pooling for
  speaker recognition,'' \emph{ArXiv}, vol. abs/2008.01077, 2020.

\bibitem{maccif}
F.~Wang, Z.~Song, H.~Jiang, and B.~Xu, ``Maccif-tdnn: Multi aspect aggregation
  of channel and context interdependence features in tdnn-based speaker
  verification,'' \emph{2021 IEEE Automatic Speech Recognition and
  Understanding Workshop (ASRU)}, pp. 214--219, 2021.

\bibitem{mfconformer}
Y.~Zhang, Z.~Lv, H.~Wu, S.~Zhang, P.~Hu, Z.~Wu, H.~yi~Lee, and H.~M. Meng,
  ``Mfa-conformer: Multi-scale feature aggregation conformer for automatic
  speaker verification,'' in \emph{INTERSPEECH}, 2022.

\bibitem{transformer}
A.~Vaswani, N.~M. Shazeer, N.~Parmar, J.~Uszkoreit, L.~Jones, A.~N. Gomez,
  L.~Kaiser, and I.~Polosukhin, ``Attention is all you need,'' in \emph{NIPS},
  2017.

\bibitem{conformer_real}
A.~Gulati, J.~Qin, C.-C. Chiu, N.~Parmar, Y.~Zhang, J.~Yu, W.~Han, S.~Wang,
  Z.~Zhang, Y.~Wu, and R.~Pang, ``Conformer: Convolution-augmented transformer
  for speech recognition,'' \emph{ArXiv}, vol. abs/2005.08100, 2020.

\bibitem{transinspeech}
P.~Guo, F.~Boyer, X.~Chang, T.~Hayashi, Y.~Higuchi, H.~Inaguma, N.~Kamo, C.~Li,
  D.~Garcia-Romero, J.~Shi, J.~Shi, S.~Watanabe, K.~Wei, W.~Zhang, and
  Y.~Zhang, ``Recent developments on espnet toolkit boosted by conformer,''
  \emph{ICASSP 2021 - 2021 IEEE International Conference on Acoustics, Speech
  and Signal Processing (ICASSP)}, pp. 5874--5878, 2021.

\bibitem{hybrid3}
F.~Xie, D.~Zhang, and C.~Liu, ``Global–local self-attention based transformer
  for speaker verification,'' \emph{Applied Sciences}, 2022.

\bibitem{finegrain1}
A.~Hajavi and A.~Etemad, ``Fine-grained early frequency attention for deep
  speaker recognition,'' \emph{2022 International Joint Conference on Neural
  Networks (IJCNN)}, pp. 1--6, 2022.

\bibitem{layercam}
P.-T. Jiang, C.-B. Zhang, Q.~Hou, M.-M. Cheng, and Y.~Wei, ``Layercam:
  Exploring hierarchical class activation maps for localization,'' \emph{IEEE
  Transactions on Image Processing}, vol.~30, pp. 5875--5888, 2021.

\bibitem{layercam2}
P.~Li, L.~Li, A.~Hamdulla, and D.~Wang, ``Reliable visualization for deep
  speaker recognition,'' in \emph{Interspeech}, 2022.

\bibitem{conformer}
Z.~Peng, W.~Huang, S.~Gu, L.~Xie, Y.~Wang, J.~Jiao, and Q.~Ye, ``Conformer:
  Local features coupling global representations for visual recognition,'' in
  \emph{2021 IEEE/CVF International Conference on Computer Vision (ICCV)},
  2021, pp. 357--366.

\bibitem{mfa}
T.~Liu, R.~K. Das, K.-A. Lee, and H.~Li, ``Mfa: Tdnn with multi-scale
  frequency-channel attention for text-independent speaker verification with
  short utterances,'' \emph{ICASSP 2022 - 2022 IEEE International Conference on
  Acoustics, Speech and Signal Processing (ICASSP)}, pp. 7517--7521, 2022.

\bibitem{se}
J.~Hu, L.~Shen, and G.~Sun, ``Squeeze-and-excitation networks,'' \emph{2018
  IEEE/CVF Conference on Computer Vision and Pattern Recognition}, pp.
  7132--7141, 2018.

\bibitem{att1}
Y.~Jiang, Y.~Song, I.~Mcloughlin, Z.~Gao, and L.~Dai, ``An effective deep
  embedding learning architecture for speaker verification,'' in
  \emph{INTERSPEECH}, 2019.

\bibitem{att2}
T.~Liu, R.~K. Das, M.~C. Madhavi, S.~syun Shen, and H.~Li, ``Speaker-utterance
  dual attention for speaker and utterance verification,'' in
  \emph{INTERSPEECH}, 2020.

\bibitem{res2net}
S.~Gao, M.-M. Cheng, K.~Zhao, X.~Zhang, M.-H. Yang, and P.~H.~S. Torr,
  ``Res2net: A new multi-scale backbone architecture,'' \emph{IEEE Transactions
  on Pattern Analysis and Machine Intelligence}, vol.~43, pp. 652--662, 2019.

\bibitem{vox2}
J.~S. Chung, A.~Nagrani, and A.~Zisserman, ``Voxceleb2: Deep speaker
  recognition,'' in \emph{Interspeech}, 2018.

\bibitem{rir}
T.~Ko, V.~Peddinti, D.~Povey, M.~L. Seltzer, and S.~Khudanpur, ``A study on
  data augmentation of reverberant speech for robust speech recognition,''
  \emph{2017 IEEE International Conference on Acoustics, Speech and Signal
  Processing (ICASSP)}, pp. 5220--5224, 2017.

\bibitem{musan}
D.~Snyder, G.~Chen, and D.~Povey, ``Musan: A music, speech, and noise corpus,''
  \emph{ArXiv}, vol. abs/1510.08484, 2015.

\bibitem{aam}
Z.~Li, Y.~Liu, L.~Li, and Q.~Hong, ``Additive phoneme-aware margin softmax loss
  for language recognition,'' \emph{ArXiv}, vol. abs/2106.12851, 2021.

\bibitem{cyc}
L.~N. Smith, ``Cyclical learning rates for training neural networks,''
  \emph{2017 IEEE Winter Conference on Applications of Computer Vision (WACV)},
  pp. 464--472, 2015.

\bibitem{vox1}
A.~Nagrani, J.~S. Chung, and A.~Zisserman, ``Voxceleb: A large-scale speaker
  identification dataset,'' in \emph{Interspeech}, 2017.

\bibitem{voxsrc}
J.~Huh, A.~Brown, J.~weon Jung, J.~S. Chung, A.~Nagrani, D.~Garcia-Romero, and
  A.~Zisserman, ``Voxsrc 2022: The fourth voxceleb speaker recognition
  challenge,'' \emph{ArXiv}, vol. abs/2302.10248, 2023.

\bibitem{asnorm}
P.~Matejka, O.~Novotn{\'y}, O.~Plchot, L.~Burget, M.~D. S{\'a}nchez, and J.~H.
  ernock{\'y}, ``Analysis of score normalization in multilingual speaker
  recognition,'' in \emph{Interspeech}, 2017.

\end{thebibliography}

\end{document}